# Robots-Assisted Redeployment in Wireless Sensor Networks

Hanen Idoudi[#1], Chiraz Houaidia[#2], Leila Azouz Saidane[#3], Pascale Minet [4]

[#]*National School of Computer Sciences, University of Manouba*
*Campus Universitaire de la Manouba, Tunisia*
[1,3]{hanen.idoudi,Leila.saidane}@ensi.rnu.tn
[2]houaidia.chiraz@gmail.com

[4] *Inria, Rocquencourt*
*78153 Le Chesnay cedex, France*
pascale.minet@inria.fr

***Abstract*** — *Connectivity and coverage are two crucial problems for wireless sensor networks. Several studies have focused on proposing solutions for improving and adjusting the initial deployment of a wireless sensor network to meet these two criteria.*
*In our work, we propose a new hierarchical architecture for sensor networks that facilitates the gathering of redundancy information of the topology. Several mobile robots must then relocate, in an optimized way, redundant sensors to achieve optimal connectivity and coverage of the network. Mobile robots have to cooperate and coordinate their movement. A performance evaluation is conducted to study the trade-off between the number of required robots and its impact on the rate of network connectivity and coverage.*

## 1. Introduction

Sensor networks are composed of small autonomous entities with low computational capabilities and limited energy resources. In addition to their basic functions (measurement, data collection, communication), sensors are often required to perform network maintenance tasks, routing and topology control. In the case of initial random deployment of a sensor network, adjusting the topology to ensure connectivity and coverage of the network must be guaranteed and may require the redeployment of sensors. One relevant solution to this problem is to provide sensors with the mobility capability so they can relocate themselves. However, all these control tasks increased by the mobility capability lead undoubtedly to excessive consumption of sensors energy which can quickly result in the complete depletion of sensors power, thus, exacerbate the lack of coverage and the connectivity problems.

Sensors and Actuators Networks are a recent development aiming at constructing heterogeneous networks composed of sensors and several other entities that may have more important processing capabilities and more energy resources (called actors or actuators). These actuators are frequently intended to assist the sensor network in order to increase its performance and extend its lifetime.

In our work, we propose, in case of faulty deployment, that actuators, mobile robots for instance, intervene to adjust the topology by redeploying sensors, hence alleviating them from both the topology maintenance and the auto-relocation tasks. At this end, we propose a new hierarchical model for sensor networks. This model aims to facilitate the collection of redundancy information by centralizing it in particular nodes called *Island-Heads*. They must communicate this information to mobile robots that scan the surface to fix connectivity and coverage holes. In addition, robots must coordinate their movement and cooperate during operation. Various simulations have allowed us to analyse the trade-off between the number of robots to use and the quality of connectivity and coverage achieved under several topologies and multiple parameters.

After a brief overview of existing researches related to our work, we discuss in Section III our new approach. Section IV is devoted to presenting the results of our simulations and their analysis. We will finish this paper with a conclusion and a review of some perspectives.

## 2. Related work

In this section, we review the most relevant existent propositions to adjust sensor network topologies. Redeployment strategies can be divided into two classes depending on the motion capabilities of sensors. In case of mobile sensors, they have to coordinate their movement to relocate themselves efficiently. In case of static sensors, actuators have to intervene to relocate sensors.

### 2.1. Redeployment schemes with mobile sensors

Several propositions rely on motion capabilities of sensors to enhance network connectivity and coverage [2, 3].

A primary matter was to model the coordinated movement of nodes and the relocation conditions to obtain a specified deployment scheme. In such schemes, sensors (or a subset of them) form a self-reconfigurable network and can move dynamically in order to adjust the topology according to the monitoring needs over the target area.

Two principal approaches are used to decide on the subset of sensors to relocate. In the first one, only redundant nodes have to move directly to the area to cover while in the second, coordinated movement of sets of nodes is proposed. In [9] the sensors are modeled as particles of a compressible fluid, in [10] the theory of gas is used to model sensor movements in the presence of obstacles. A similar approach is used in [11] to give a unified solution to the problem of deployment and dynamic relocation of mobile sensors in an open environment. The Voronoi approach is used in [12], where mobile sensors move from densely deployed areas to sparse areas on the basis of a local calculation of the Voronoi diagrams. In other solutions, [13] propose the use of Delaunay triangulation techniques to obtain a regular tessellation of the area of interest.

However, these solutions are costly in terms of consumed energy and delays since redundant sensors may have to travel long distances to improve the coverage. To balance the energy cost and the replacement time, a cascading movement is proposed in [5]. Once a coverage hole is detected, nodes move towards it along a selected path. Algorithms related to potential field and virtual forces that relocate nodes are presented in [7]. Areas of redundancy or high density are reduced thanks to the repulsive forces exerted between neighbouring nodes. Redundant nodes are moved to heal coverage holes.

### 2.2. Redeployment schemes in Wireless Sensor and Actuators Networks

Nevertheless, mobility feature in WSN and particularly in large-scale sensor networks is costly in terms of power consumption of sensors. An attractive alternative is the use of actuators to assist redeployment of sensors, which gave birth to the Wireless Sensor and Actuators Networks (WSAN).

Indeed, actors, robots for instance, can be used to perform maintenance tasks on static sensor networks. Mobile robots can assist the redeployment of redundant sensors to sparse areas. Although using robots to assist deployment or maintenance of sensor networks is an attractive solution, few studies have focused on proposing new schemes in such context.

Authors propose in [1] the use of some mobile robots to assist the replacement of exhausted sensors or recharge their batteries. All the robots are mobile and can take and drop sensors in pre-calculated positions. When a node becomes inoperative due to its energy's depletion, a robot moves to the target position, replaces the faulty sensor by a functional one or recharges the battery of the exhausted sensor. In a first centralized manager algorithm, a central manager receives failure reports from sensors and forwards them to individual robots. However in the distributed version, each robot functions as both a manager and a maintainer. Nevertheless, in this work, authors do not precise how to locate and collect redundant sensors. Authors consider in [4] one mobile robot to assist the deployment of static sensor networks. The proposed deployment scheme uses permanent grid and cluster concepts to reduce the number of packets used in creating and maintaining a grid structure. The proposed solution aims at reducing the robot's motion and efficiently guides the robot to redeploy sensors. However, a grid-based architecture is not feasible where nodes are relatively randomly deployed. In such case, the cost of re-organizing sensors into grids is high. Moreover, redeploying sensors in a large field with a single robot is challenging. Indeed, algorithms here described take potentially several iterations to terminate. They may not meet the requirements of a fast redeployment.

### 3. Robots-Driven sensor network redeployment

We present in this section our approach to redeploy the sensor network using mobile robots. We begin with the network modeling, and the proposed clustering scheme based on this model. We detail next the robots functioning during the network redeployment and the robots cooperation protocol proposed for this purpose.

### 3.1. Network Modelling and Hierarchy

In this work we assume that the field of interest is hardly accessible which makes quite impossible any deterministic initial deployment. Therefore, we consider initially a random deployment which consists on spreading a large amount of sensors on the field from, for instance, an airplane. Following such deployment, several isolated sets of nodes are formed called here *islets*. We call *islet (or Island)* each set of connected nodes unable to reach the sink. The *Mainland* is the islet including the sink node. Connectivity and coverage within an islet are ensured through the high density and redundancy of nodes. Figure 1 describes our Islet-based model.

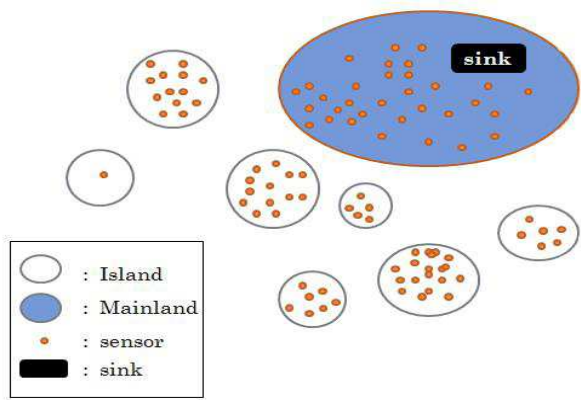

Figure 1 Islets-based topology

To facilitate the location of redundant sensors, we use a hexagonal partitioning of the field such that any two sensors in two adjacent cells can still communicate. A cell is covered as long as it includes at least one sensor. A sensor node is considered redundant if its perception zone (represented here by a cell) is already covered by other nodes.

Based on this network modelling, we propose a clustering scheme that aims to aggregate information related to each islet and reduce the energy consumption (see Figure 2). This algorithm consists on considering each islet as a cluster. For each cell, a *Master* node should be elected and remains in the active state to provide coverage within the cell and gather information about redundant nodes. Other nodes of the same cell are considered redundant and go into the sleeping mode. Among all the *Master* nodes in an islet, an *Island-Head* is chosen, according to its residual energy and its position within the islet, to be the local coordinator within. *Island-Heads* collect information about the number, positions and the energy of redundant nodes on each islet and should provide robots with this information when needed. Further details on both *Master* nodes and *Island-Heads* election mechanisms were previously discussed [17].

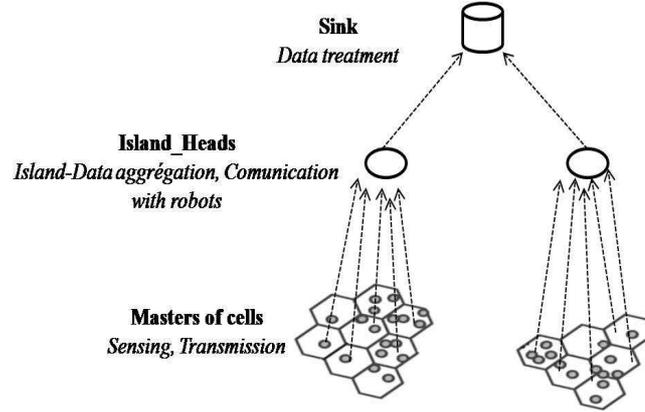

Figure 2 Nodes Hierarchy

As an improvement to this scheme, we propose to rotate the role of *Master* and *Island-Head* between different sensors. Indeed, nodes having these roles are required to be in a continuous active mode which can deplete their energy. To balance the energy consumption equitably among all nodes, we propose to re-elect periodically Master and Island-Heads.

### 3.2. Topology enhancement

Robots can carry a certain number of sensors that could be used to heal connectivity and coverage holes. They carry an initial load of sensors and they are able to collect redundant sensors as they find them, as long as their maximum load is not reached.

Furthermore, robots have to coordinate their movement and synchronize their functioning. Robots perform topology discovery and heal connectivity and coverage holes simultaneously.

We divide the surface into zones which width is less than the transmission range of robots. Each robot has to move across one or several assigned zones as depicted by Figure 3.

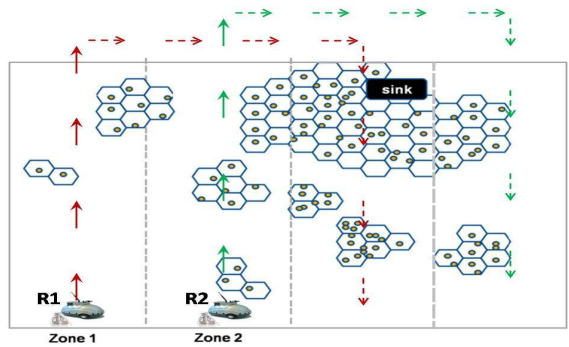

Figure 3 Robots Mobility Pattern

When crossing a zone, a robot has to discover the topology. It has to broadcast a *HELLO_Robot* message periodically. If an *Island-Head* intercepts this message, it has to respond to the robot by indicating the number and location of all redundant sensors within its islet. If a robot does not receive any response to its *HELLO_Robot* message after a certain travelled distance, it considers the travelled zone as a coverage hole and puts a sensor to heal it. This functioning is described by the following algorithm (see Figure 4). If a robot finds redundant sensors, it can retrieve and carry them in order to place them when needed. This is done unless its maximum capacity of carried sensors is not reached. In this case, the robot has to memorize redundant sensors locations in order to retrieve them in the next iteration.

When reaching one of the two horizontal edges of the surface, a robot has to synchronize with its neighbouring robots. It stops to communicate with them in order to lend them some sensors, if they request it, or to ask them for some sensors. A robot requests for additional sensors from its neighbours if the carried sensors are below a critical threshold. Its neighbours can lend it if they have enough carried sensors on their own i.e. more than the specified threshold. This mechanism is explained by the algorithm in Figure 5.

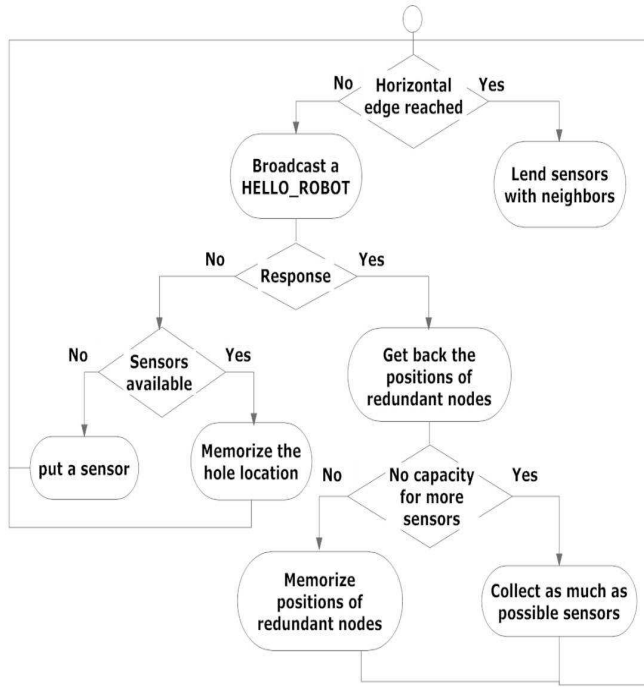

Figure 4 Robots functioning

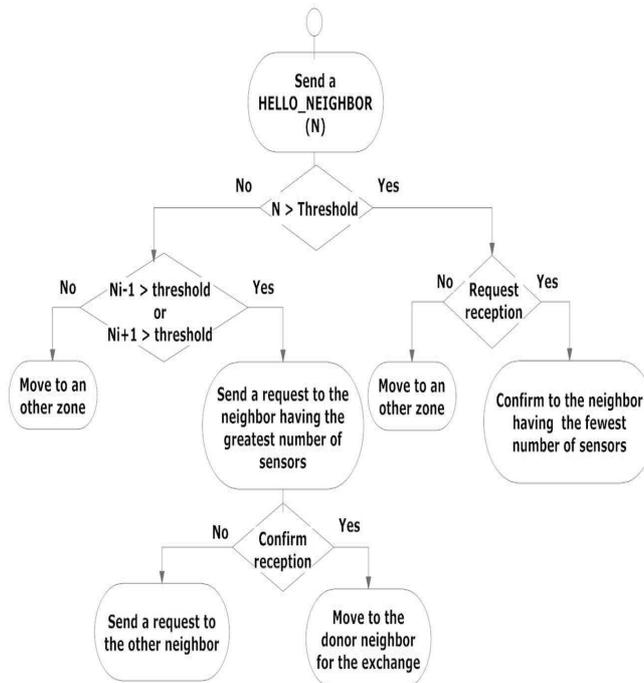

Figure 5 Robots Cooperation

Robots may have to undergo several iterations to cover all connectivity and coverage holes. In case of a second crossing of the field and more, the same behaviour will be reproduced except for communication with sensors. Indeed, the robots no longer need to collect information on the number and position of the redundant nodes from the Island-heads as this information has been stored since the first passage. The robots can then proceed directly to the redundant sensors and carry them in order to place them when needed, while maintaining the same model of coordinated mobility and cooperation between robots.
The elimination of communication with the Island-Heads in the following passages is intended to reduce the overhead, accelerate the robots functioning and conserve Island-Heads energy.

4. **Performance Evaluation**

We implemented our proposed approach under the WSNet simulator [15]. Different scenarios were established to allow a detailed performance assessment of our solution. In each simulation, we checked that the total number of available sensors is quite sufficient to ensure full coverage of the surface. We carried our simulations while varying both the number of robots and the initial topology.

We varied the number of robots in order to determine the optimal number of robots required to ensure connectivity and network coverage. On the other hand, we varied the initial network topology (the number of islets) in order to evaluate the impact of redundancy on coverage and connectivity.

The following table summarizes the different simulation parameters we used.

| Parameter | Value |
|---|---|
| Network dimensions ($L \times H$) | 600*600 m$^2$ |
| Side's length of a cell (*Hexa_cote*) | 7.5 m |
| Number of sensors (*N*) | 600 |
| Sensor communication radii (*Rc*) | 30 m |
| Sensor sensing radii (*Rs*) | 15 m |
| Initial load of a robot | 60 sensors |
| Robot's communication range (*R*) | 17,5 m |
| Robot's threshold of sensors (*Threshold*) | 5 |

TABLE I. SIMULATION PARAMETERS

Redundancy is a key parameter on performance evaluation of our solution. A primary impact of the islet-based model, that we adopted, is the trade off we notice between the number of created islets and the redundancy. As we noticed by Figure 6, redundancy decreases when distributing sensors on several islets.

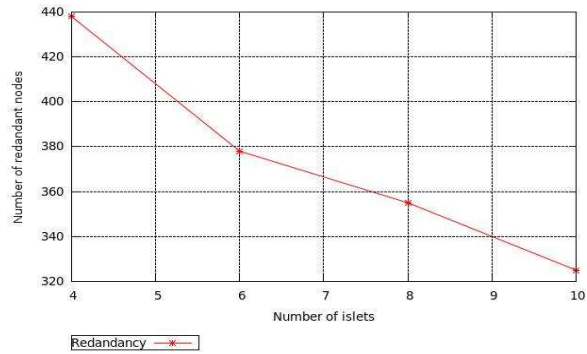

Figure 6 Redundancy Rating

### 4.1. Coverage ratio

According to the hexagonal pattern, we estimate the surface coverage by considering a cell as the coverage unit. A cell is covered as long as it contains at least one sensor. Coverage ratio (COR) is thus determined by the number of covered cells and is defined as follows:

$$COR = number\ of\ covered\ cells\ /\ total\ number\ of\ cells \qquad (1)$$

Figure 7 shows the coverage ratio at the end of one field's pass when using 2, 4 or 6 robots. It follows that with more robots, we have automatically better coverage results within a single surface crossing. We point out also that the more the number of islets increases, the less the coverage ratio is. This is explained by the fact that redundancy decreases when sensors are scattered into multiple islets.

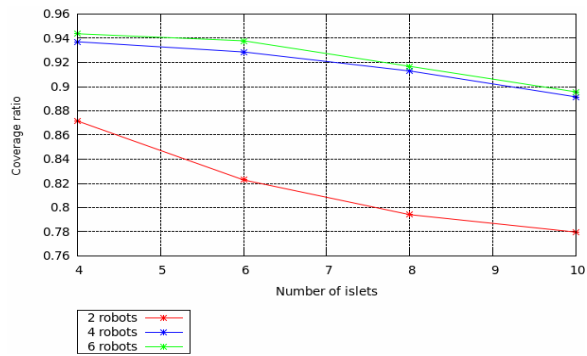

Figure 7 First Round Coverage Ratio

Furthermore, if we consider Figure 8, we notice the very important enhancement of the coverage ratio for all subsequent topologies in comparison to the ratios of the initial topologies (up to 90% for a 4 islets topology, for instance). Under such improvement, we can easily predict that a total coverage could be reached within few more passes of the field.

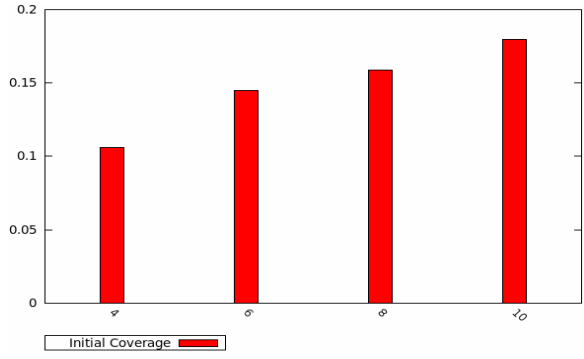

Figure 8 Initial Coverage Ratio

### 4.2. Connectivity ratio

Connectivity is a vital feature to a sensor network since it expresses the network's ability to route data to the sink node.

We measured the connectivity ratio by considering the number of islets connected to the mainland. At this end, we define the connectivity ratio in terms of islets (CRI). This number, as mentioned by equation 2, corresponds to the difference between the initial number of islets and the final number of still isolated islets.

$$CRI = (initial\ number\ of\ islets - final\ number\ of\ islets) / initial\ number\ of\ islets \qquad (2)$$

As expected, simulations show that increasing the number of robots improves the connectivity ratio reached at the end of the first round (see Figure 9). Moreover, using at least 2 robots allows a good connectivity ratio (up to 80%) in case of a 10 initial islets topology. Furthermore, as depicted in Figure 9, we notice that the connectivity ratio increases while using topologies with higher number of islets. This is explained by the fact that in initial topologies with a high number of islets, sensors are more scattered through the field than in few islets topologies. Hence, islets are separated by few cells and total connectivity could be reached more quickly by placing few sensors to connect islets.

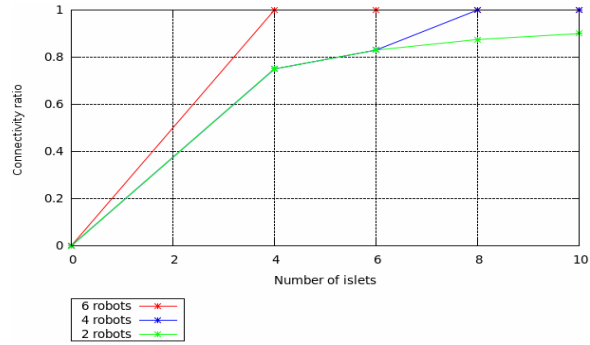

Figure 9 First Round Connectivity Ratio

On the other hand, we varied the size of islets in order to study the connectivity of sensors to the sink. We point out that the connectivity ratio grows differently according to the topology to achieve progressively important connectivity reaching 100% (see Figure 10).

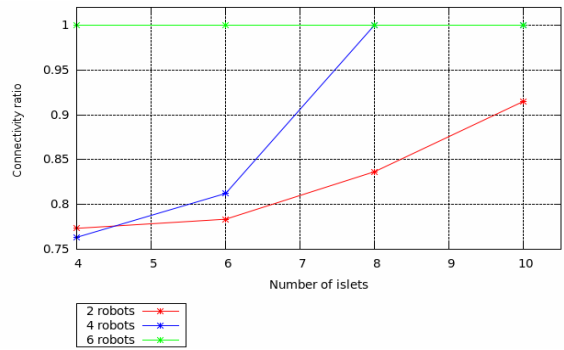

Figure 10 Connectivity Ratio in Terms of Sensors

### 4.3 Coverage time

In a second series of experiments, we measure the time needed for robots to perform full connectivity and coverage of the network. Thus, we can study the trade-off between the number of robots used and the time required for their intervention. This time depends on three steps: the surface exploration including the communication with the *Island-Heads*, the collection of redundant sensors and robot's synchronization for cooperation.

Figure 11 demonstrates the evolution of coverage time while varying the number of robots and the initial topology. The time represented includes all rounds performed by robots to ensure full coverage. Generally, only one pass is enough for some topologies to achieve full connectivity but to complete network coverage, we observe that at least two iterations are required to. The second pass is restricted to the collection of redundant nodes and their relocation.

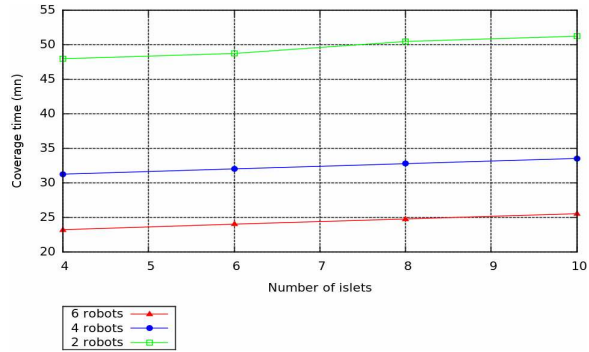

Figure 11 Coverage Time

As shown in Figure 11, two robots take significantly more time to perform the redeployment than four or six robots. On the other hand, the coverage time increases slightly with the number of islets which can be explained by the increase, of time to collect redundant nodes and communication delays between robots and sensors (the "island heads" in particular).

**4.4 Energy consumption**

The high energy consumption driven by sensor's mobility is a major criterion which justifies the use of static sensors and the use of actuators to redeploy them. Moreover, in our solution, several mechanisms are proposed in the perspective of reducing the energy consumption of sensors.

Mainly, the clustering scheme and the sleeping mode for the redundant nodes should contribute significantly to save sensors power. In addition, data aggregation by the *Island-Heads* should lead to the reduction of the number of packets exchanged with the sink and with robots and thus the energy dissipated to do so.

The impact of our proposal on sensors energy is studied to highlight the trade-off between the cost induced by the use of robots and the effectiveness of the proposed approach towards coverage and connectivity objectives.

Figure 12 represents the mean consumed energy according to the topology. The effect of redundancy is confirmed through this representation. In fact, we note that energy consumption increases with the number of islets which is due to the smaller number of sleeping sensors with several islets.

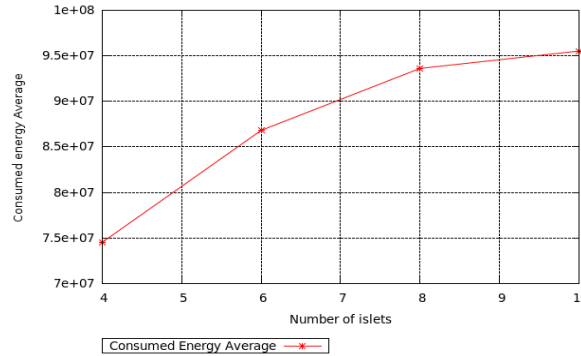

Figure 12 Average Consumed Energy

**5. Conclusion and future work**

Sensors have been increasingly adopted in the context of several disciplines and applications (military, industrial, medical, homeland security, etc.) with the aim of collecting and distributing observations of a target field. Using robots to assist these networks is an emergent paradigm which aims at avoiding human intervention on sensor networks deployed in hazardous fields or accelerating their maintenance.

In this paper we proposed a new scheme for sensor networks redeployment assisted by mobile robots. Our proposal aims at improving the coverage and the connectivity of the monitored area while using minimum number of mobile robots. The main idea of the proposed redeployment is to exploit the redundancy induced by the initial random deployment in order to overcome connectivity and coverage holes. The redundant sensors are put into sleeping mode and will be relocated by robots into uncovered cells following an optimal placement. Robots perform the network redeployment in a cooperative way while exploring the sensor network.

Simulation results showed the efficiency of our solution in terms of both coverage and connectivity enhancement. Results have proven that using only few mobile robots has an important impact on reaching a total connectivity of the network while significantly enhancing the coverage ratio after a single robots crossing of the monitored area.

In our future work, we intend to consider other mobility strategies for robots that could optimize the time for achieving connectivity and coverage. We also intend, in order to maintain the coverage within the area, to assist robots to systematic monitoring for detecting and replacing faulty sensors.